\documentclass[longauth,structabstract]{aa}  
\usepackage{graphicx}
\usepackage{natbib}
\usepackage{color} 
\usepackage{txfonts}
\bibpunct{(}{)}{;}{a}{}{,} 
\newcommand{\nc}{\newcommand}

\newcommand{\CII}{[C\,{\sc ii}]}
\newcommand{\CI}{[C\,{\sc i}]} 
\newcommand{\SII}{[S\,{\sc ii}]}
\newcommand{\NII}{[N\,{\sc ii}]}
\newcommand{\NIII}{[N\,{\sc iii}]}
\newcommand{\OIII}{[O\,{\sc iii}]}

\newcommand{\SIII}{[S\,{\sc iii}]}
\newcommand{\OI}{[O\,{\sc i}]}

\newcommand{\HII}{H\,{\sc ii}}
\newcommand{\HI}{H\,{\sc i}}
\newcommand{\SIV}{[S\,{\sc iv}]}
\newcommand{\NeII}{[Ne\,{\sc ii}]}
\newcommand{\Nethree}{[Ne\,{\sc iii}]}
\newcommand{\Halpha}{H$\alpha$}
\nc\micron{\mbox{$\mu$m}}
\nc{\cmcub}{\mbox{cm$^{-3}$}}
\nc{\cmsq}{\mbox{cm$^{-2}$}}
\nc{\Kkms}{\mbox{K~km~s$^{-1}$}}
\nc{\kms}{\mbox{km~s$^{-1}$}}
\nc{\mthirty}{\mbox{M\,33}}
\nc{\Tmb}{\mbox{$T_{\rm mb}$}}
\nc{\vlsr}{\mbox{v$_{\rm LSR}$}}
\newcommand\arcdeg{\mbox{$^\circ$}}%
\nc{\twCO}{$^{12}$CO}
\nc{\thCO}{$^{13}$CO}
\nc{\msun}{\ensuremath{\mathrm{M}_\odot}}
\nc{\rsun}{\ensuremath{\mathrm{R}_\odot}}
\nc{\lsun}{\ensuremath{\mathrm{L}_\odot}}
\newcommand\hour{\mbox{$^{\mathrm h}$}}%
\newcommand\minute{\mbox{$^{\mathrm m}$}}%

\begin{document}
\title{The Herschel M33 extended survey (HerM33es): PACS spectroscopy
  of the star forming region BCLMP\,302 \thanks{Herschel is an ESA
    space observatory with science instruments provided by
    European-led Principal Investigator consortia and with important
    participation from NASA.}}


\titlerunning{PACS spectroscopy of BCLMP\,302}

   \author{B. Mookerjea \inst{\ref{tifr}}, 
   C. Kramer\inst{\ref{iramsp}},
   C. Buchbender\inst{\ref{iramsp}}, 
   M. Boquien \inst{\ref{umass}},
   S. Verley \inst{\ref{ugr}},
   M. Rela\~{n}o\inst{\ref{ugr}},
   G. Quintana-Lacaci\inst{\ref{iramsp}}, 
   S. Aalto \inst{\ref{onsala}},
   J. Braine \inst{\ref{bordeaux}},
   D. Calzetti \inst{\ref{umass}},
   F. Combes \inst{\ref{obspm}},
   S. Garcia-Burillo \inst{\ref{oanmadrid}},
   P. Gratier \inst{\ref{bordeaux}},
   C. Henkel \inst{\ref{mpifr}},
   F. Israel \inst{\ref{leiden}},
   S. Lord,\inst{\ref{ipac}},
   T. Nikola \inst{\ref{cornell}},
   M. R\"ollig \inst{\ref{kosma}},
   G. Stacey \inst{\ref{cornell}},
   F. S. Tabatabaei \inst{\ref{mpifr}},
   F. van der Tak \inst{\ref{sron}},
   P. van der Werf \inst{\ref{leiden},\ref{supa}}
          }
\institute{Tata Institute of Fundamental Research, Homi Bhabha Road,
Mumbai 400005, India \email{bhaswati@tifr.res.in}\label{tifr}
\and
Instituto Radioastronom\'{i}a Milim\'{e}trica, Av. Divina Pastora 7,
Nucleo Central, E-18012 Granada, Spain \label{iramsp}
\and
University of Massachusetts, Department of Astronomy, LGRT-B 619E, 
Amherst, MA 01003, USA \label{umass}
\and
Universidad de Granada, E-18012 Granada, Spain \label{ugr}
\and
Department of Radio and Space Science, Onsala Observatory,
Chalmers University of Technology, S-43992 Onsala, Sweden
\label{onsala}
\and
Laboratoire d'Astrophysique de Bordeaux, Universit\'{e} Bordeaux
1, Observatoire de Bordeaux, OASU, UMR 5804, CNRS/INSU, B.P.
89, Floirac F-33270 \label{bordeaux}
\and
Observatoire de Paris, LERMA, CNRS, 61 Av. de l�Observatoire,
75014 Paris, France \label{obspm}
\and 
Observatorio Astron\'{o}mico Nacional (OAN) - Observatorio de
Madrid, Alfonso XII 3, 28014 Madrid, Spain \label{oanmadrid}
\and
Max Planck Institut f\"ur Radioastronomie, Auf dem H\"ugel 69, D-
53121 Bonn, Germany \label{mpifr}
\and
Leiden Observatory, Leiden University, PO Box 9513, NL 2300
RA Leiden, The Netherlands \label{leiden}
\and
IPAC, MS 100-22 California Institute of Technology, Pasadena, CA
91125, USA \label{ipac}
\and
Department of Astronomy, Cornell University, Ithaca, NY 14853 \label{cornell}
\and
KOSMA, I. Physikalisches Institut, Universit\"at zu K\"oln,
Z\"ulpicher Strasse 77, D-50937 K\"oln, Germany \label{kosma}
\and
Institute of Astronomy, University of Cambridge, Madingley Road,
Cambridge CB3 0HA, England \label{cambridge}
\and
SRON Netherlands Institute for Space Research, Landleven 12, 9747 AD
Groningen, The Netherlands 19  \label{sron}
\and
SUPA, Institute for Astronomy, University of Edinburgh, Royal
Observatory, Blackford Hill, Edinburgh EH9 3HJ, UK 7 \label{supa}
             }

 \date{Received \ldots; accepted \ldots}

\authorrunning{Mookerjea et al.}








\abstract
{The emission line of \CII\ at $158\,\mu$m is one of
    the strongest cooling lines of the interstellar medium (ISM) in
    galaxies.}
  {Disentangling the relative contributions of the different ISM
    phases to \CII\ emission, is a major topic of the HerM33es
    program, a Herschel key project to study the ISM in the nearby
    spiral galaxy M33.  }
  {Using PACS, we have mapped the emission of \CII\ 158$\,\mu$m, \OI\
63$\,\mu$m, and other FIR lines in a 2\arcmin$\times$2\arcmin\ region
of the northern spiral arm of M33, centered on the \HII\ region
BCLMP\,302.  At the peak of H$\alpha$ emission, we have observed in
addition a velocity resolved \CII\ spectrum using HIFI.  We use
scatterplots to compare these data with PACS 160$\,\mu$m\ continuum
maps, and with maps of CO and \HI\ data, at a common resolution of
12\arcsec\ or 50\,pc.  Maps of \Halpha\ and 24$\,\mu$m\ emission
observed with Spitzer are used to estimate the SFR.  We have created
maps of the \CII\ and \OI\ 63$\,\mu$m\ emission and detected \NII\
122$\,\mu$m\ and \NIII\ 57$\,\mu$m\ at individual positions.  }
  {The \CII\ line observed with HIFI is significantly broader than that
    of CO, and slightly blue-shifted. In addition, there is little
    spatial correlation between \CII\ observed with PACS and CO over
    the mapped region. There is even less spatial correlation between
    \CII\ and the atomic gas traced by \HI. Detailed comparison of the
    observed intensities towards the \HII\ region with models of photo
    ionization and photon dominated regions, confirms that a
    significant fraction, 20--30\%, of the observed \CII\ emission stems
    from the ionized gas and not from the molecular cloud. 
    The gas heating efficiency, using the ratio between \CII\ and the 
    TIR as a proxy, varies between 0.07 and 1.5\%, with the largest 
    variations found outside the \HII\ region.  }
{}


\maketitle

%

\section{Introduction}

\begin{figure}[h]
\centering
\includegraphics[height=9cm,angle=-90]{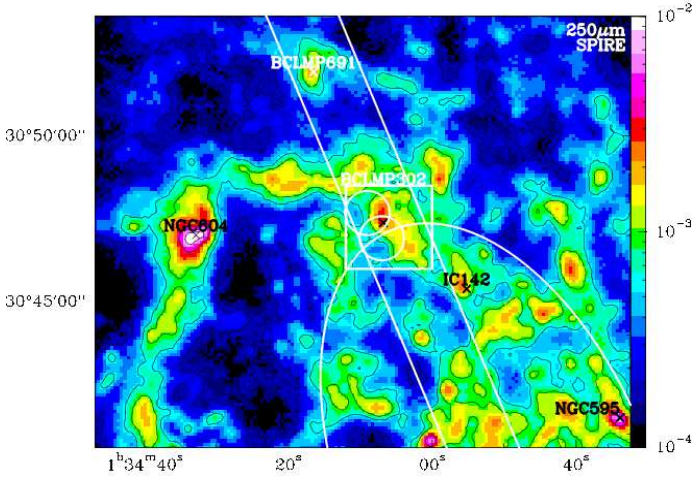}
\caption{250\,$\mu$m Herschel-SPIRE dust continuum image of the
northern spiral arm of M33 along with the locations of the prominent
\HII\ regions NGC\,604, NGC\,595, IC\,142, BCLMP\,302, and
BCLMP\,691. The rectangle delineates the area observed with PACS and
HIFI, which is centered on BCLMP\,302. The white circles indicate the
positions and $70''$ beam of ISO/LWS observations done within the
BCLMP\,302 region. The white ellipse delineates 2\,kpc galacto centric
distance. The two parallel lines running along the major axis of M33,
mark the strip for which we plan to observe the \CII\ and other
FIR lines using PACS and HIFI.
\label{fig_overview}}
\end{figure}

\begin{figure}[h]
\centering
\includegraphics[width=9cm,angle=0]{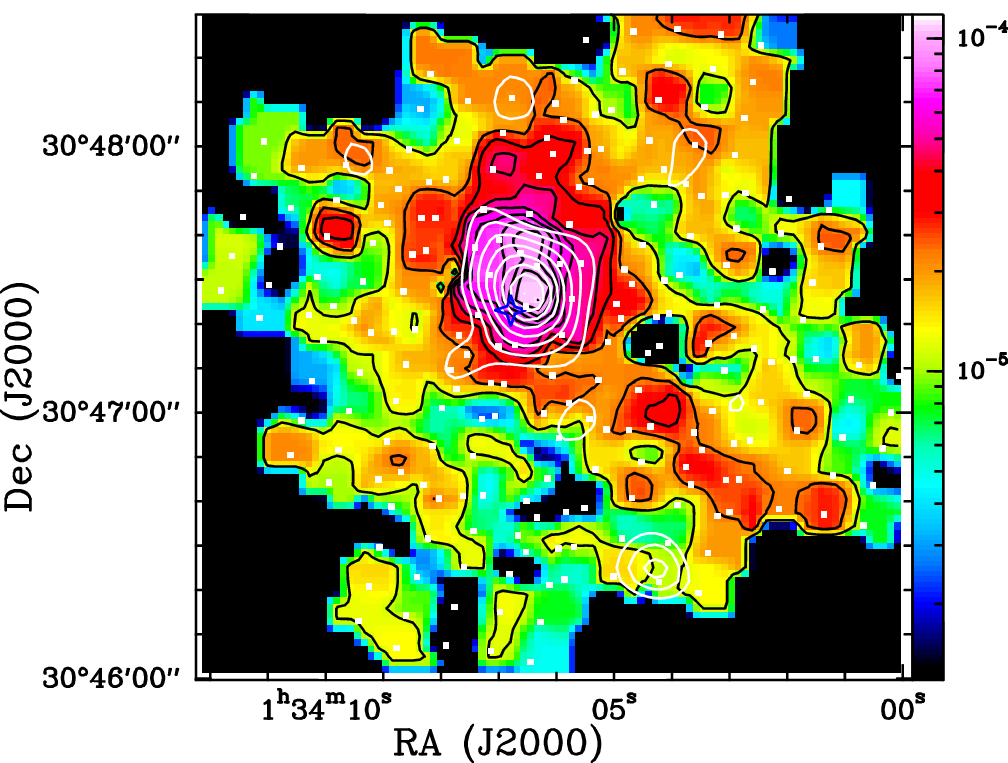}
\caption{Maps of 158$\,\mu$m\ \CII\ (in color and black contours) and
  63$\,\mu$m\ \OI\ (white contours) emission observed with PACS toward
  BCLMP\,302. The \CII\ intensities shown in the color wedge are in
  units of erg\,s$^{-1}$\,cm$^2$\,sr$^{-1}$. The \OI\ 63$\,\mu$m\ map
  has been smoothed to 12\arcsec\ for easy comparison with \CII. The
  \Halpha\ peak observed with HIFI is marked with the asterisk. The
  white dots show the footprint of the PACS observations. The contour
  levels are at 10 to 100\% (in steps of 10\%) of peak \CII\ intensity
  of 1.18$\times 10^{-4}$\,erg\,cm$^{-2}$\,s$^{-1}$\,sr$^{-1}$. The
  contour levels for \OI\ 63$\,\mu$m\ emission are 30 to 100\% (in
  steps of 10\%) of the peak of 3.0$\times
    10^{-5}$\,erg\,cm$^{-2}$\,s$^{-1}$\,sr$^{-1}$.  Both images are at
    a common resolution of 12\arcsec.
    \label{fig_ciiresult}}
\end{figure}

\begin{figure}[h]
\centering
\includegraphics[width=9cm]{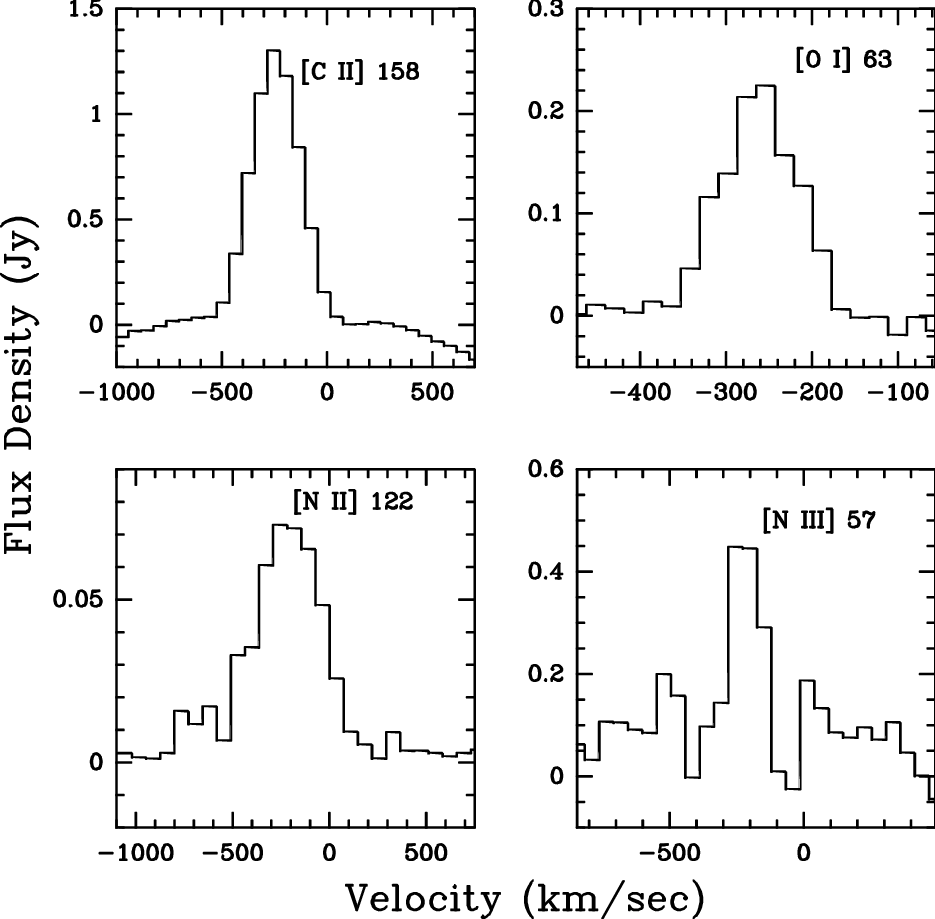}
\caption{PACS spectra of \CII\ (158~\micron), \OI\ (63~\micron), \NII\
(122~\micron) and \NIII\ 57~\micron\ at the \Halpha\ peak
position of BCLMP\,302. The LSR velocity is given.
All lines are unresolved, i.e. line profiles only reflect 
the instrumental profiles.
\label{fig_pacsspec}}
\end{figure}

\begin{figure}[h]
\begin{center}
\includegraphics[width=7.0cm]{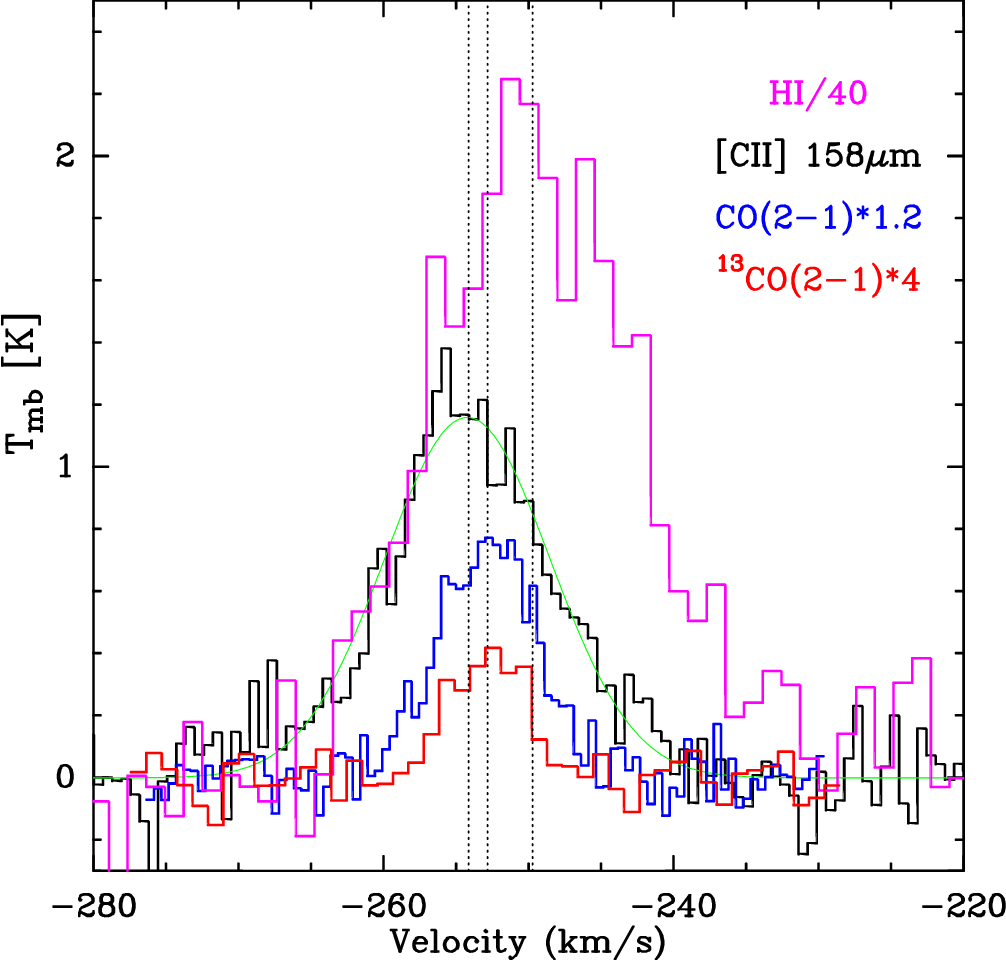}
\caption{All four spectra of \CII, \HI\ and the (2--1) transition
of CO and \thCO\ at the H$\alpha$ peak position of the \HII\ region
BCLMP 302. All four spectra are at $\sim12''$ resolution allowing
for a detailed comparison. The vertical lines denote the velocities
-249.7, -252.8, -254.1~\kms\ corresponding to \HI, CO, and \CII\
emission respectively. 
\label{fig_allspec}}
\end{center}
\end{figure}

\begin{figure*}[h!] \centering
\includegraphics[width=11.0cm,
angle=-90]{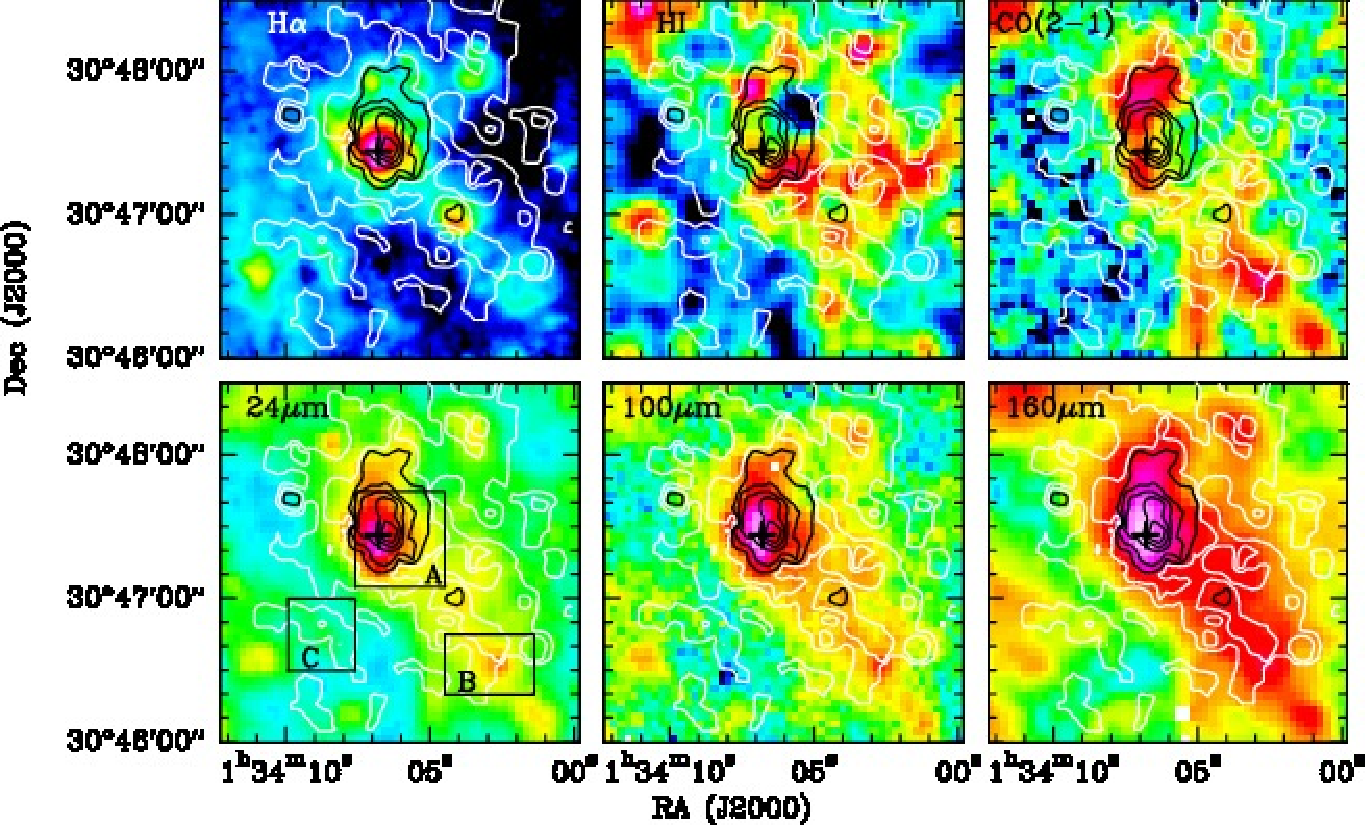} 
\caption{Overlay of the \Halpha, \HI\ and \twCO(2--1) emission and the
  dust continuum at 24 (MIPS), 100 \& 160$\,\mu$m\ (PACS) emission
  with contours of \CII\ at 158$\,\mu$m. The white and black contours
  are for intensities between 10--20\% (in steps of 10\%) and 30--90\%
  (in steps of 15\%) of the peak \CII\ intensity of 1.18$\times
  10^{-4}$~erg~s$^{-1}$~cm$^{-2}$~sr$^{-1}$.  The beam sizes for
  \Halpha, \HI, \twCO(2--1), 24$\,\mu$m, PACS 100$\,\mu$m, PACS
  160$\,\mu$m\ and \twCO(2--1) observations are 1\arcsec, 12\arcsec,
  12\arcsec, 6\arcsec, 6\farcs7 and 11\farcs4 respectively. The \HI\
  map is integrated between -280 km/s to -130 \kms.  and the CO(2--1)
  map is integrated between -270 to -220~\kms. 
   Marked in the 24$\,\mu$m\ image (with black
  rectangles) are also the regions selected for further analysis.
\label{fig_allmaps}}
\end{figure*}

The thermal balance and dynamics of the interstellar medium in
galaxies, is best studied through spectroscopic observations of its
major cooling lines: \CI, \CII, and \OI\ trace the transition regions
between the atomic and molecular gas, while CO traces the dense
molecular gas that provides the reservoir for stars to form.  Most of
the important cooling lines lie in the far-infrared and submillimeter
regime.  Therefore, it is difficult or impossible to study them with
ground-based telescopes, while previous space-based telescopes
provided low sensitivity and coarse angular resolution.  The infrared
line emission is mostly optically thin and can be traced throughout
the densest regions in galaxies, allowing an unhindered view of the
ISM.  Herschel provides, for the first time, an opportunity to image
the major tracers of the ISM at a sensitivity, spectral and spatial
resolution that allows to study the interplay between star formation
and the active ISM throughout our Milky Way and in nearby galaxies.

The two fine structure lines of \CII\ at 158$\,\mu$m and \OI\ at
63\,$\mu$m, are the strongest cooling lines of the ISM, carrying up to
a few percent of the total energy emitted from galaxies in the
far-infrared wavelengths.  The \CII\ line lies 92\,K above the ground
state with a critical density for collisions with H of
$3\times10^3$\,cm$^{-3}$ \citep{Kaufman1999}.  While both lines are
thought to trace photon-dominated regions (PDRs) at the FUV irradiated
surfaces of molecular clouds, it has been realized early-on that a
non-negligible fraction of the \CII\ emission may stem from the
ionized medium \citep{Heiles1994}. Owing to its higher upper energy
level (228\,K) and higher critical density of $\simeq 5\times
10^5$\,\cmcub, the \OI\ 63$\,\mu$m\ line is a more dominant coolant in
warmer and denser neutral regions \citep[e.g.][]{Roellig2006}.

The \CII\ line together with the \OI\ line are diagnostics to infer
the physical conditions in the gas, its temperatures, densities, and
radiation fields, by comparing the intensities and their ratios with
predictions of PDR models \citep[e.g.,][]{TielensHollenbach1985,
Wolfire1990, Kaufman1999, Roellig2007, Ferland1998}. Previous
observational studies of \CII\ emission from external galaxies include
the statistical studies by
\citet{Crawford1985,stacey1991,malhotra2001}.  More recently the \CII\
emission from a few individual galaxies e.g., LMC \citep{israel1996},
M51 \citep{nikola2001,kramer2005}, NGC~6946
\citep{Madden1993,Contursi2002}, M83 \citep{kramer2005}, NGC~1313
\citep{Contursi2002}, M~31 \citep{rodriguez2006}, NGC1097
\citep{Contursi2002,Beirao2010} have been observed. These papers explore
the origin of the \CII\ emission. Of these studies,
the study of LMC by \citet{israel1996} was at a resolution of 16~pc
and \citet{rodriguez2006} resolved the spiral arms of M~31 at spatial
scales of 300~pc.

In addition to \CII\ and \OI\ 63\micron, there are several additional
FIR lines, the mid-$J$ CO transitions, and lines of single and double
ionized N and O which provide information about the gas density
(multiple transitions of the same ions), hardness of the stellar
radiation field (ratio of intensities of two ionization states of the
same species), and ionizing flux. In particular the \NII\ lines at 122
and 205$\,\mu$m\ allow us to estimate the density of the ionized gas,
which is a key parameter to model the \CII\ emission stemming from the
ionized gas.

Although much detailed information can be obtained by studying nearby
(Milky Way) sites of star formation, a more comprehensive view is
possible with a nearby moderately inclined galaxy such as M33. M\,33
is a nearby, moderately metal poor late-type spiral galaxy with no
bulge or ring, classified as SA(s)cd. It is the 3rd largest member of
the Local Group. Its mass, size, and average metallicity are similar
to those of the Large Magellanic Cloud (LMC). M33 hosts some of the
brightest \HII\ complexes in the Local Group.  NGC\,604 is the second
brightest \HII\ region after 30\,Doradus in the LMC.  Its inclination
($i=56^\circ$) \citep{Regan1994} yields a small line-of-sight depth
which allows to study individual cloud complexes not suffering from
distance ambiguities and confusion like Galactic observations do.  Its
close distance of 840\,kpc \citep{Freedman1991} provides a spatial
resolution of 50\,pc at $12''$, allowing us to resolve giant molecular
associations in its disk with current single dish millimeter and
far-infrared telescopes like the IRAM 30m telescope and Herschel.  Its
recent star formation activity, together with the absence of signs of
recent mergers, makes M\,33 an ideal source to study the interplay of
gas, dust, and star formation in its disk.  This is the aim of the
open time key program ``Herschel M33 extended survey'' {\sf HerM33es}
\citep{kramer2010}. To this end we are surveying the major cooling
lines, notably \CII, \OI, \NII, as well as the dust spectral energy
distribution (SED) using all three instruments on board of Herschel,
HIFI, PACS, and SPIRE. We also use ancillary observations of \Halpha,
\HI, CO and dust continuum at 24$\,\mu$m.  The {\tt
    HerM33es} PACS and HIFI spectral line observations will focus on a
  number of individual regions along the major axis of M~33 which will
  initially be presented individually. 
%

Here, we present first spectroscopic results obtained for BCLMP302, one
of the brightest \HII\ regions of M33. We discuss a $2'\times2'$
($\sim$0.5\,kpc $\times$ 0.5\,kpc) field in the northern spiral arm at a
galactocentric distance of 2\,kpc (Figure\,\ref{fig_overview}). This
region lies along the major axis of M33, and houses one of the bright
\HII\ regions, BCLMP\,302 \citep{boulesteix1974, israel1974}. Using
ISO/SWS, \citet{wilner2002} studied Neon abundances of \HII\ regions in
M33, including BCLMP\,302.  \citet{rubin2008} used {\it Spitzer}-IRS to
map the emission lines of \SIV\ 10.51, H(7--6) 12.37, \NeII\ 12.8,
\Nethree\ 15.56 and \SIII\ 18.71$\,\mu$m in 25 \HII\ regions in M33,
including BCLMP\,302.  ISO/LWS \CII\ unresolved spectra at $\sim70''$
resolution \citep{gry2003} are available for this region from the
archive. Here, we present PACS maps of \CII\ and \OI\ 63$\,\mu$m at a
resolution of 12\arcsec, together with a HIFI \CII\ spectrum at 2\,\kms\
velocity resolution.  We compare these data with (i) the \Halpha\
emission tracing the ionized gas, (ii) dust continuum images at mid- and
far-infrared wavelengths observed with {\it Spitzer} and Herschel,
tracing the dust heated by newly formed stars and the diffuse
interstellar radiation field, and (iii) CO and \HI\ emission tracing the
neutral molecular and atomic gas.

The rest of the paper is organized as follows: Sec. 2 presents details
of our observations and ancillary data, Sec. 3 states the basic
results of the spectroscopic observations and a qualitative and a
quantitative comparison of the \CII\ and \OI\ emission with all other
available tracers and their correlation. 
%
Sec.~4 studies the role of \CII\ as an indicator of the star formation
rate (SFR) and Sec.~5 analyzes the energy balance in the mapped
region.  Sec.~6 presents a detailed analysis of the emission from the
\Halpha\ peak position in BCLMP\,302 in terms of models of PDRs. In
Sec.~7 we summarize and discuss the major findings of the paper.


\section{Observations}

\subsection{Herschel: PACS Mapping}

A region extending over 2\arcmin$\times$2\arcmin\ around the \HII\
region BCLMP 302 in the northern arm of M\,33 was observed with the
5$\times$5 pixel integral field unit (IFU) of the PACS Spectrometer
using the wavelength switching (WS) mode in
  combination with observations of an emission free off source
  position outside of the galaxy at RA/Dec (J2000) =
  22.5871$^\circ$/30.6404$^\circ$.  The field of view of the IFU is
47\arcsec$\times$47\arcsec\ with 9\farcs4 pixels
\citep{poglitsch2010}.  We used the 1st and 3rd order gratings to
observe the \CII\ 157.7$\,\mu$m, \OI\ 63.18$\,\mu$m, \OI\
145.52$\,\mu$m, \NII\ 121.9$\,\mu$m, \NIII\ 57.3$\,\mu$m, and \NII\
205.18$\,\mu$m\ lines, with the shortest possible observing time (1
line repetition, 1 cycle), and with a reference position at RA =
01$^{\rm h}$30$^{\rm m}$20\fs9, Dec = 30\arcdeg 38\arcmin 25\farcs4
(J2000). The reference position was selected based on \HI\ and
100\micron\ ISSA\footnote{IRAS Sky Survey Atlas} maps.

For all the lines a 3$\times$3 raster was observed on a 40\arcsec\ grid
with the IFU centered at R.A.  = 01$^{h}$ 34$^{m}$ 05\fs9 Dec=+30\arcdeg
47\arcmin 18\farcs6 (J2000) with position angle, P.A. = 22.5\arcdeg.
The resulting footprint is shown in Fig.\,\ref{fig_ciiresult}.  The FWHM
beam size of the PACS spectrometer is 9.2\arcsec\ near 63$\,\mu$m\ and
11.2\arcsec\ near 158$\,\mu$m\ (E.Sturm, priv. comm.). The lines are
unresolved, as the spectral resolution of PACS is larger than 90\,\kms\
for all lines. 
The observations were performed on January 7, 2010 and the total
observing time was 1.1\,hours for all the 6 lines.  The PACS spectra
were reduced using HIPE version 3.0 CIB 1452 \citep{ott2010}.  The WS
data reduction pipeline was custom-made by the NASA Herschel Science
Center (NHSC) helpdesk. The data were exported to FITS cubes which were
later analyzed using internally developed IDL routines to extract the
line intensity maps.

Using PACS, the \CII 158$\,\mu$m, \OI\ 63$\,\mu$m, \NII\ 122$\,\mu$m,
and the \NIII\ 57$\,\mu$m\ lines\ have been detected . The lines of
\OI\ 145$\,\mu$m\ and the \NII\ 205$\,\mu$m\ were not detected.  The
peak and $1\sigma$ noise limits of the intensities of the \CII, \OI,
\NII\ 122$\,\mu$m\ and \NIII\ 57$\,\mu$m\ spectra, at the \Halpha\
peak position, are presented in Table~\ref{tab_linerms}.
Fig.~\ref{fig_pacsspec} show the observed PACS spectra at the
position of the \Halpha\ peak position. The peak integrated
intensities were derived by first fitting and subtracting a polynomial
baseline of 2nd order and then fitting a Gaussian.  Due to unequal
coverage of different positions, as seen from the grid of observed
positions shown in Fig.\,\ref{fig_ciiresult}, the rms achieved is not
uniform. It varies by about a factor of 3 over the entire map.

\begin{table}[h]
  \caption{Peak, signal to noise ratio (SNR), and sigma values
of the integrated line intensities at the \Halpha\ peak 
position for the lines detected with PACS all at original resolution. 
\label{tab_linerms}}
\begin{tabular}{lrr}
\hline
\hline
Line & Peak (SNR) & $1\sigma$ \\
& erg~s$^{-1}$~cm$^{-2}$~sr$^{-1}$ & erg~s$^{-1}$~cm$^{-2}$~sr$^{-1}$\\
\hline
\CII\ 158$\,\mu$m & 1.18$\times 10^{-4}$ (67) & 1.72$\times 10^{-6}$\\
\OI\ 63$\,\mu$m & 7.20$\times 10^{-5}$ (6) & 1.20$\times 10^{-5}$\\
\NII\ 122$\,\mu$m & 9.95$\times 10^{-6}$ (7) & 1.42$\times 10^{-6}$\\
\NIII\ 57$\,\mu$m & 1.30$\times 10^{-5}$ (7) & 1.80$\times 10^{-6}$ \\
\hline
\end{tabular}
\end{table}

\subsection{Herschel: HIFI spectrum at the  \Halpha\ peak}

Using HIFI we have observed a single spectrum at the peak position (R.A.
= 01$^{\rm h}$34$^{\rm m}$06\fs79 Dec = 30\arcdeg 47\arcmin 23\farcs1
(J2000)) of the \Halpha\ emission from BCLMP\,302
(Fig.\,\ref{fig_allspec}).  The HIFI spectrum was taken on 01 August
2010 during one hour of observing time using the load chop mode with the
same reference position as was used for the PACS observation.  The
frequency of the \CII\ line is 1900536.9~MHz, known to within an
uncertainty of 1.3~MHz (0.2~\kms) \citep{Cooksy1986}.  The blue shifted
line required to tune the local oscillator to 1899.268 GHz, about the
highest frequency accessible to HIFI.  The \CII\ spectra were recorded
using the wide band acousto optical spectrometer, covering a bandwidth
of 2.4\,GHz for each polarization with a spectral resolution of 1\,MHz.
We calculated the noise-weighted averaged spectrum, combining both
polarizations. A fringe fitting tool available within HIPE was used to
subtract standing waves, subsequently the data were exported to CLASS
for further analysis.  Next, a linear baseline was subtracted and the
spectrum was rebinned to a velocity resolution of 0.63~\kms\
(Fig.~\ref{fig_allspec}).  We scaled the resulting data to the main beam
scale using a beam efficiency of 69\%, using the Ruze formula with the
beam efficiency for a perfect primary mirror $\eta_{{\rm mb},0}=0.76$
and a surface accuracy of $\sigma=3.8\,\mu$m \citep{OlbergBeam2010}.
The measured peak temperature from a Gaussian fit is 1.14\,K and the rms
($1\sigma$ limit) is 110\,mK at 0.63\,kms$^{-1}$ resolution, which is
consistent within 10\% of the rms predicted by HSPOT. The half power
beam width (HPBW) is $12.2''$.


\subsection{IRAM 30m CO observations}

For comparison with the \CII\ spectrum we have observed spectra of the
(2--1) and (1--0) transitions of \twCO\ and \thCO, at the position of
the \Halpha\ peak, using the IRAM 30\,m telescope on 21 August 2010.
These observations used the backend VESPA. The spectra were smoothed
to a velocity resolution of 1~\kms\ for all the CO lines. The forward
and beam efficiencies are 95\% and 80\% respectively for the (1--0)
transition. The same quantities are 90\% and 58\% for the (2--1)
transition. The half-power beam widths for the (1--0) and (2--1)
transitions are 22\arcsec\ and 12\arcsec, respectively.

\subsection{Complementary data for comparison}

We compare the Herschel data of the BCLMP\,302 region, with the \HI\ VLA
and CO(2--1) HERA/30m map at $12''$ resolution presented by
\citet{gratier2010}. We refer to the latter paper for a presentation of
the noise properties. 

We also use the {\it Spitzer} MIPS 24\,\micron\ map presented by
\citet{tabatabaei2007}, the KPNO \Halpha\ map \citep{hoopes2000}, and
the {\tt HerM33es} PACS and SPIRE maps at 100, 160, 250, 350 and
500\,\micron\ \citep{kramer2010, verley2010, boquien2010}. The angular
resolution of the 100 and 160~\micron\ PACS maps are $\sim 6$\arcsec\
and 12\arcsec.  The rms noise levels of the PACS maps are
2.6\,mJy~pix$^{-2}$ at 100$\,\mu$m\ and 6.9\,mJy~pix$^{-1}$ at
160$\,\mu$m\ where the pixel sizes are 3\farcs2 and 6\farcs4
respectively.

\CII\ observations at two positions within our mapped region were
extracted from the ISO/LWS archival data. The two positions are at RA=
01$^{\rm h}$34$^{\rm m}$07$^{\rm s}$ Dec=30\arcdeg 46\arcmin
55\arcsec\ (J2000) and RA=01$^{\rm h}$34$^{\rm m}$09$^{\rm s}$
DEC=30\arcdeg47\arcmin41\arcsec\ (J2000).

\section{Results}

\subsection{PACS Far-Infrared Spectroscopy}


Within the $2'\times2'$ region mapped with PACS
(Fig.\,\ref{fig_ciiresult}), we have detected extended \CII\ emission
from (a) the northern spiral arm traced e.g. by the $100\,\mu$m
emission, with the strongest emission arising towards the \HII\ region
BCLMP\,302 and (b) from the diffuse regions to the south-east and
north-west.  Comparison of the \CII\ PACS intensities with the ISO/LWS
\CII\ data at the two positions shows an agreement of better than 11\%
at both positions.  For the comparison of the intensities at the LWS
positions, we first convolved the PACS \CII\ map to the angular
resolution of the LWS data. The full ISO/LWS \CII\ data set along the
major axis of M33 will be published in a separate paper by Abreu et al.
(in prep.).

Some of the PACS \OI\ 63$\,\mu$m\ spectra displayed baseline problems,
attributed to the now decommissioned wavelength switching mode.
Spectra taken along the \CII\ emitting part of the spiral arm
extending north-east to south-west showed problems and have been
blanked.  Both, the \NII\ 122$\,\mu$m\ and the \NIII\
57$\,\mu$m\ lines were detected at only a few positions within the
mapped region. The maps of \OI\ 63$\,\mu$m\ and \CII\ 158$\,\mu$m\
emission towards the \HII\ region look very similar and both peak at
RA = 01$^{\rm h}$34$^{\rm m}$ 06\fs3, Dec =
30\arcdeg47\arcmin25\farcs30. In addition, the \OI\ 63$\,\mu$m\ map
shows a secondary peak towards the south-west, to the south of the
\CII\ ridge, at RA=01\hour34\minute04\fs364 Dec =
30\arcdeg46\arcmin26\farcs55 (J2000), which is not found in the \CII\
map. The second \OI\ peak lies between the two ridges detected
in CO(2--1) and coincides with an \HI\ peak. This suggests that the
\OI\ emission at this peak position arises in very dense atomic gas.

Overlays of the \CII\ map with maps of \Halpha, \HI, CO(2--1) and dust
continuum in the MIR and FIR (MIPS 24$\,\mu$m, PACS 100 \&
160$\,\mu$m) are shown in Fig.~\ref{fig_allmaps}.  The dust continuum
maps correlate well with the \CII\ map. They peak towards the \HII\
region and show the spiral arm extending from the \HII\ region in
south-western direction. In contrast, CO emission shows a clumpy
structure wrapping around the \HII\ region towards the east. CO
emission shows the spiral arm seen in the continuum, but its peaks are
shifted towards the south. The \HI\ emission shows a completely
different morphology, peaking towards the north and south of the \HII\
region and showing a clumpy filament running towards the west. Further
below, we will discuss the correlations in more detail.

\begin{figure*}[h]
\centering
\includegraphics[height=16.0cm, angle=-90]{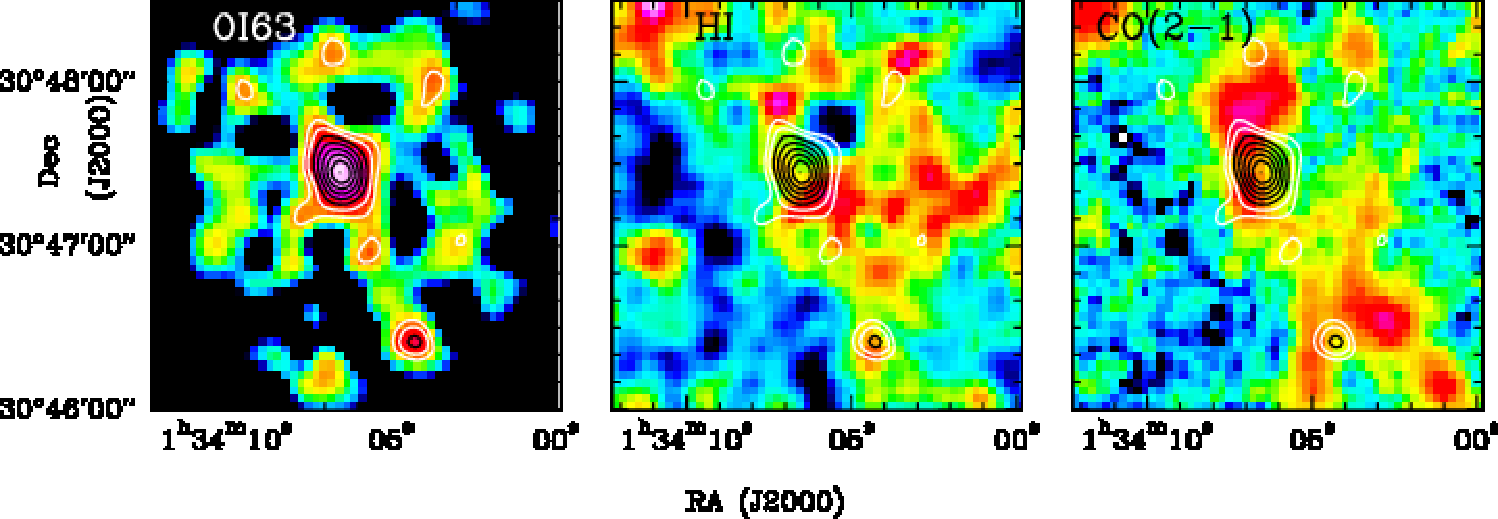}
\caption{Overlay of color plots of \OI\ 63$\,\mu$m, \HI\ and
\twCO(2--1) emission with contours of \OI\ at 63$\,\mu$m. {\bf All plots
are at a resolution of 12\arcsec. }
The white and black contours are for intensities
between 30--40\% (in steps of 10\%) and 50--100\% (in steps of 10\%)
of peak \OI\ intensity of 3.0$\times
10^{-5}$~erg~s$^{-1}$~cm$^{-2}$~sr$^{-1}$. Details of the \HI\ and
\twCO(2--1) maps are identical to those in Fig.~\ref{fig_allmaps}.
\label{fig_o63olay}}
\end{figure*}

Fig.~\ref{fig_o63olay} shows overlays of the \OI\ 63$\,\mu$m\ map at a
resolution of 12\arcsec\ with \HI\ and CO(2--1).  Towards the south
and south-west of the \HII\ region, the \OI\ 63$\,\mu$m\ emission
matches the \HI\ emission well. This is surprising given the high
excitation requirements for the \OI\ line. The secondary \OI\
63$\,\mu$m\ peak towards the south-west, is also traced by \HI. It
lies in between two ridges of CO emission, the arm running north-east
to south-west, and a second ridge of emission running in north-south
direction. The northern part of this second ridge shows an interesting
layering of emission: both \OI\ and \HI\ are slightly shifted towards
the east relative to this CO ridge. However, there is much less
correspondence between \OI\ and \HI\ emission towards the east and
north of the \HII\ region. 

\subsection{HIFI spectroscopy of the \Halpha\ region}

Fig.~\ref{fig_allspec} shows the velocity resolved 158$\,\mu$m\ \CII\
spectrum observed with HIFI at the H$\alpha$ peak position. In
addition, we show the spectra of \HI\ and the J=2--1 transitions of CO
and \thCO.  HIFI and PACS integrated intensities agree very well. The
\CII\ integrated intensity of 15.6~\Kkms, observed with HIFI
corresponds to an intensity of
1.10$\times$10$^{-4}$~erg~s$^{-1}$~cm$^{-2}$~sr$^{-1}$ and this
matches extremely well with the \CII\ intensity observed with PACS, at
the nearest PACS position, which is offset by only 3\arcsec.
%


All spectra are at a common resolution of $\sim12$\arcsec, allowing
for a detailed comparison. All spectra show a Gaussian line shape.
However, the line widths are strikingly different between the atomic
material traced by \HI, the \CII\ line, and the molecular gas.  The
\HI\ spectrum shows a FWHM of 16.5~\kms, that of \CII\ is 13.3~\kms,
while CO 2--1 shows a width of only 8~\kms
(Table~\ref{tab_gaussprof}). We take this as an indication that the
\HI\ disk along the line of sight is much thicker than the molecular
disk while the material traced by \CII\ appears to lie in between.  In
addition, we find that the lines are not centered at the same
velocity. The HI line is shifted by $+4.4$~\kms\ relative to \CII,
while the CO lines are shifted by $+1.6$~\kms\ relative to \CII,
confirming that all three tracers trace different components of the
ISM. The shifts are significant, as the error of the Doppler
corrections are much smaller, for HIFI (D.Teyssier, priv.  comm.) as
for the other data.

Based on lower spectral resolution (6~\kms) optical spectroscopy of
H$\beta$ and \OIII\ emission lines, the velocity of the ionized gas is
deduced to be shifted by $-16.9$~\kms\ with respect to the \CII\ line,
with a velocity dispersion of 11~\kms\ \citep{wilner2002,zaritsky1989}.
The optical spectroscopic data is however at much higher angular
resolution (2--4\arcsec). A map of the emission lines originating solely
from the ionized medium, which would allow to smooth these data to
12\arcsec\ resolution for direct comparison with the other data, is not
yet available.  In Section 6.2, we will use PDR models to show that
about 20--30\% of the observed \CII\ stems from the ionized gas of the
BCLMP\,302 \HII\ region.

In summary, we find a ``layering'' of the line-of-sight velocities
tracing the different ISM components. The velocity increases from
H$\alpha$ to \CII\ to CO to \HI.

\begin{table}[h]
\begin{center}
  \caption{Parameters derived from Gaussian fits to spectra observed
    with HIFI, VLA, and IRAM 30m, at the \Halpha\ peak position
    (Figs.\,\ref{fig_ciiresult} \& \ref{fig_allspec}).
    $\theta_{\rm b}$ indicates the half power beamwidth.
    \label{tab_gaussprof}}
\begin{tabular}{lrlccc}
\hline \hline
Line & $\theta_{\rm b}$ & $\int{T {\rm dv}}$ & $v_{\rm cen}$ & $\Delta v$ \\
     & $''$             & \Kkms & \kms & \kms \\
\hline
\CII\ & 11 & 15.6$\pm$0.5 & -254.1$\pm$0.1 & 12.9$\pm$0.9 \\
\HI\  & 11 & 1446.7$\pm$44.1 & -249.7$\pm$0.2 & 16.5$\pm$0.6\\
CO (1--0) & 22 & 3.2$\pm$0.2 & -252.3$\pm$0.2 & 7.5$\pm$0.5 \\
CO (2--1) & 11 & 5.3$\pm$0.1 & -252.8$\pm$0.1 & 7.9$\pm$0.3 \\
\thCO (1--0) & 22 & 0.28$\pm$0.02 & -252.9$\pm$0.2 & 5.8$\pm$0.5\\
\thCO (2--1) & 11 & 0.69$\pm$0.06 & -252.8$\pm$0.3 & 6.2$\pm$0.5\\
\hline
\end{tabular}
\end{center} 
\end{table}

\subsection{Detailed comparison of \CII\ emission with other tracers
\label{sec_scatplot}}

\begin{figure*}[ht!]
\centering
\includegraphics[width=11.0cm, angle=-90]{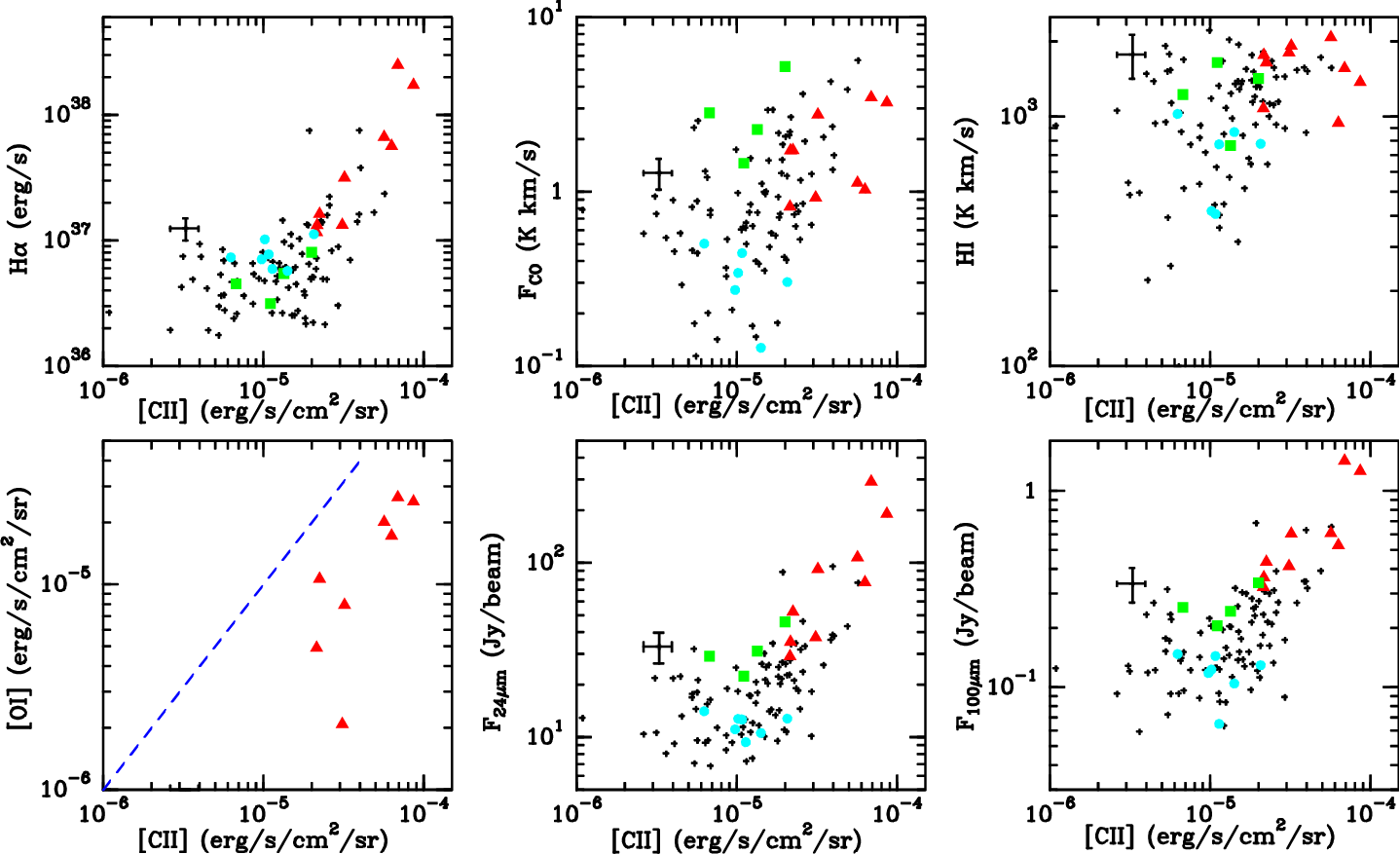}
\caption{Correlation of   \CII\ with \Halpha,  \twCO(2--1), \HI,  \OI\
  63$\,\mu$m,     MIPS   24$\,\mu$m\      and   PACS      100$\,\mu$m.
  The black crosses mark all  positions on a 6\arcsec\
    grid.  The red  triangles  correspond to  positions in region $A$
  region, the  green  squares represent  positions in  region $B$
  and the cyan filled circles correspond  to  positions in  region $C$. 
   All  these regions are marked in  the 24$\,\mu$m\ map in
  Fig.~\ref{fig_allmaps}.  Errorbars corresponding  to a 20\% error on
  the    plotted   quantities along   both   axes   are shown  at  one
  representative  point  in  each  panel.   For \OI\  63$\,\mu$m, only
  positions close to the \HII\ region are used.   The blue dashed line
  in the \CII--\OI\ scatter  plot corresponds to equal  intensities of
  the two tracers.
  \label{fig_allscats}}
\end{figure*}

\begin{table}[h]
\begin{center}
\caption{Correlation coefficients for the scatter plots in
Fig.~\ref{fig_allscats}.
\label{tab_corrcoeff}}
\begin{tabular}{lcccc}
\hline \hline
Tracers & \multicolumn{4}{c}{Correlation Coefficient}\\
& Entire & Region& Region & Region \\
& Map & $A$ & $B$ & $C$\\
\hline
\CII--\Halpha & 0.60 & 0.93 & 0.66 & 0.30 \\
\CII--CO(2--1) & 0.41 & 0.40 & 0.47 & \ldots\\
\CII--\HI & 0.22 & $<0.1$ & $<$0.10 & 0.14\\
\CII--\OI & \ldots & 0.77 & \ldots & \ldots \\
\CII--F$_{24}$ & 0.61 & 0.86 & 0.66 & 0.41\\
\CII--F$_{100}$ & 0.59 & 0.85 & 0.56 & 0.43\\
\hline
\end{tabular}
\end{center} 
\end{table}

\begin{table*}[h]
\begin{center}
\caption{Average values and their variation of the different tracers
arising from the \HII\ region, the spiral arm and the 
region $C$.
\label{tab_regionstat}}
\begin{tabular}{lccc}
\hline \hline
&&&\\
Tracer & Region $A$ & Region $B$ & Region $C$ \\
\hline
&&&\\
\CII\ (erg~s$^{-1}$~cm$^{-2}$~sr$^{-1}$) & ($4.51 \pm 2.30$)$\times 10^{-5}$ & ($1.28
\pm 0.48$)$\times 10^{-5}$ & ($1.19 \pm 0.42$)$\times 10^{-5}$ \\
\Halpha\ (erg\,s$^{-1}$) & (7.1$\pm$6.9)$\times 10^{37}$ &
(5.3$\pm$1.8)$\times 10^{36}$ & (7.95$\pm$1.93)$\times 10^{36}$ \\
CO (2--1) (K\,km\,s$^{-1}$) & 1.88$\pm$0.98 & 2.95$\pm$1.40 & 0.29$\pm$0.14 \\
\HI\ ((K\,km\,s$^{-1}$)  & 1597$\pm$358 & 1267$\pm$325 & 620$\pm$309 \\
\OI\ (erg~s$^{-1}$~cm$^{-2}$~sr$^{-1}$)& (1.28$\pm$0.09$\times 10^{-5}$ & \ldots & \ldots \\
F$_{24}$ (Jy\,beam$^{-1}$) & 101.9$\pm$82.8 & 32.2$\pm$8.6 & 11.9$\pm$1.50 \\
F$_{100}$ (Jy\,beam$^{-1}$) & 0.66$\pm$0.37 & 0.26$\pm$0.05 & 0.12$\pm$0.03 \\
&&&\\
\hline
\end{tabular}
\end{center} 
\end{table*}

For a more quantitative estimate of the correspondence between the
different tracers in which the spiral arm has been mapped we have
created scatterplots of intensities of tracers like \Halpha, \twCO,
\HI, \OI\ 63$\,\mu$m, MIPS 24$\,\mu$m\ and PACS 100$\,\mu$m\ as a
function of the \CII\ intensities (Fig.~\ref{fig_allscats}).
We have used intensities from all the maps which are smoothed
to a resolution of 12\arcsec\ and gridded on a 12\arcsec\ grid. In
order to identify any apparent trends in the emission we have defined
three sub-regions within the mapped region: Region $A$
corresponds to the \HII\ region BCLMP\,302 itself, Region $B$
corresponds to the south-western more quiescent part of the spiral
arm, traced by e.g.  the 100$\,\mu$m continuum emission, and centered
on a peak of CO emission, and Region $C$ lies outside of the
prominent CO arm.  These three sub-regions are marked by black
rectangles in Figure\,\ref{fig_allmaps}. A second \HII\ region along
the arm, lies just outside and to the north of the box defining the
quiescent arm region.  For the remainder of the paper we always
analyze and compared the results for these three sub-regions
separately.

In the log-log scatter plots of Figure\,\ref{fig_allscats}, we find that
the \Halpha, \OI, and continuum emission at 24$\,\mu$m\ and 100$\,\mu$m\
show pronounced linear correlations with the \CII\ emission within the
\HII\ region.  In the region $C$, the intensities of the
\Halpha\ emission and the continuum emission at 24 and 100$\,\mu$m\
remain almost constant. The CO(2--1) intensity in the \HII\ region
($A$) is only poorly correlated with the \CII\ emission.  In
regions $B$ and $C$ the CO(2--1) intensity shows no correlation
with the \CII\ intensity and has a large scatter.  \HI\ does not show
any correlation with \CII.

A more quantitative analysis of the correlation between the different
tracers and \CII, is obtained by calculating the correlation
coefficient ($r$) (Table~\ref{tab_corrcoeff}). For the entire region
\Halpha\ and 24$\,\mu$m\ and 100$\,\mu$m\ intensities show around 60\%
correlation with the \CII\ intensities.  For the region $A$ \Halpha,
\OI, 24$\,\mu$m, and 100$\,\mu$m\ are well correlated ($r>0.75$) with
\CII. The \Halpha\ emission is strongly correlated with the \CII\
intensity also on the south-western arm position.  
The \OI/\CII\ intensity ratio measured primarily in the \HII\ region
varies between 0.1--0.4 and this variation is significantly larger
than the estimated uncertainties. 

\section{\CII\ as a tracer of star formation
\label{sec_starform}}

\begin{figure}[h] \centering
\includegraphics[width=7.0cm]{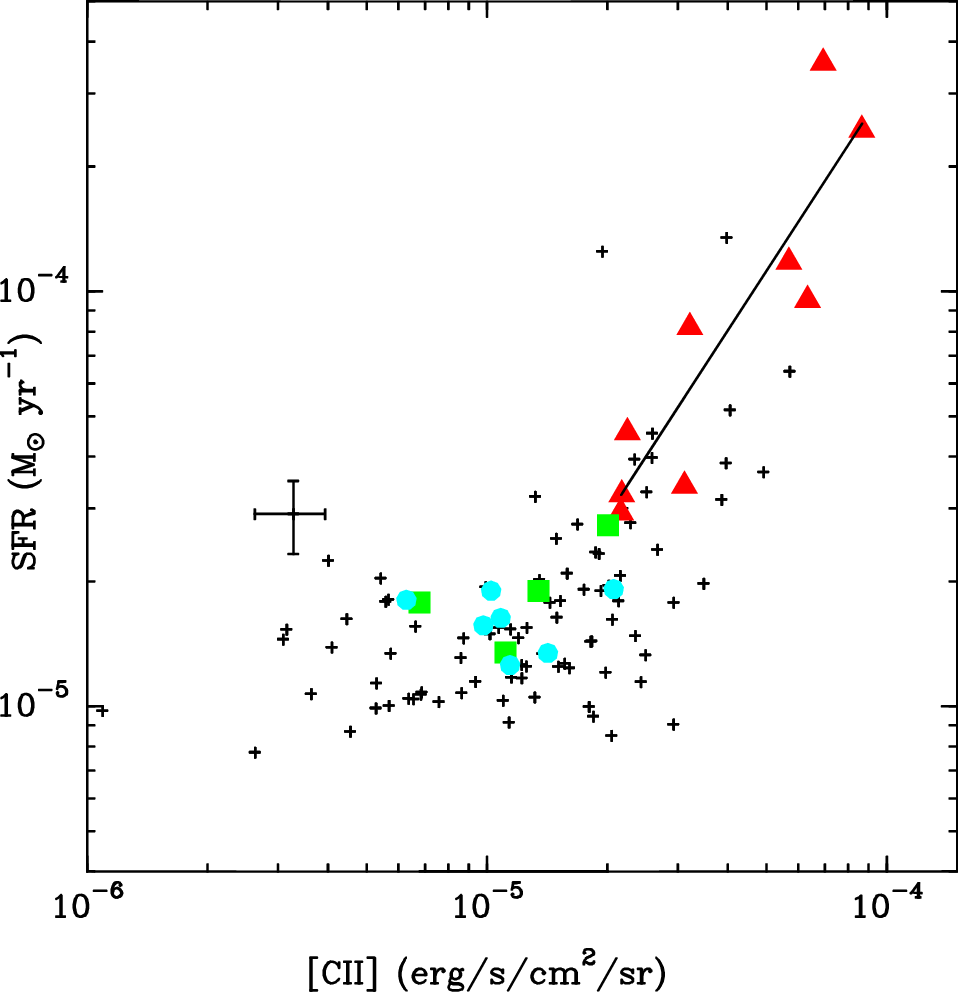}
\caption{Correlation of \CII\ intensities with SFR (derived from
  24$\,\mu$m\ + H$\alpha$).  Red  triangles  correspond to  
  positions in region $A$ region, the  green  squares represent  
  positions in  region $B$ and the cyan filled circles correspond  
  to  positions in  region $C$.  Each marker corresponds to one 
  position on
  a $12''$ grid. Details about the regions and points are identical to
  Fig.~\ref{fig_allscats}.  The black straight line marks the
  fits to the region $A$.  The errorbar corresponds to an uncertainty
  of 20\%.
\label{fig_scatplot}}
\end{figure}

In Fig.~\ref{fig_scatplot}, we plot the star formation rate (SFR),
estimated from the 24$\,\mu$m\ MIPS data and the KPNO H$\alpha$ data, as
a function of the \CII\ intensity.  Positions within the selected
regions $A$, $B$, and $C$ are marked using different symbols
and colours similar to Fig.~\ref{fig_allscats}.  The SFR has been
calculated from SFR$~=~[L(H\alpha)~+~0.031L(24)]~\times~5.35~\times
10^{-35}$ in $\msun$~yr$^{-1}$, where L(\Halpha) is the \Halpha\
luminosity in Watt and L(24) is defined as $\nu L_\nu$ at 24$\,\mu$m\ in
Watt \citep{calzetti2007}. We have assumed a \citet{kroupa2001} initial
mass function with a constant SFR over 100~Myr. Here, we study the
correlation on scales of $12''$ corresponding to 50\,pc. On these small
scales, we may start to see a break-up of any tight correlations between
the various tracers of the SFR.

Viewing at all pixels we find a steepening of the slope in the log--log
plots from regions where both the SFR and \CII\ are weak, to regions
where both are strong. Towards the \HII\ region ($A$), we find an almost
linear relation between $\log$(SFR) and $\log$(\CII), with a good
correlation, $r=0.90$ and the fitted slope is 1.48$\pm$0.26, i.e. the
SFR goes as \CII\ to the power 1.48.  Region $B$ shows a correlation
coefficient of 0.64 and region $C$ shows no correlation.

\section{Energy Balance in the spiral arm}

\begin{figure}[h]
\centering
\includegraphics[width=7.0cm]{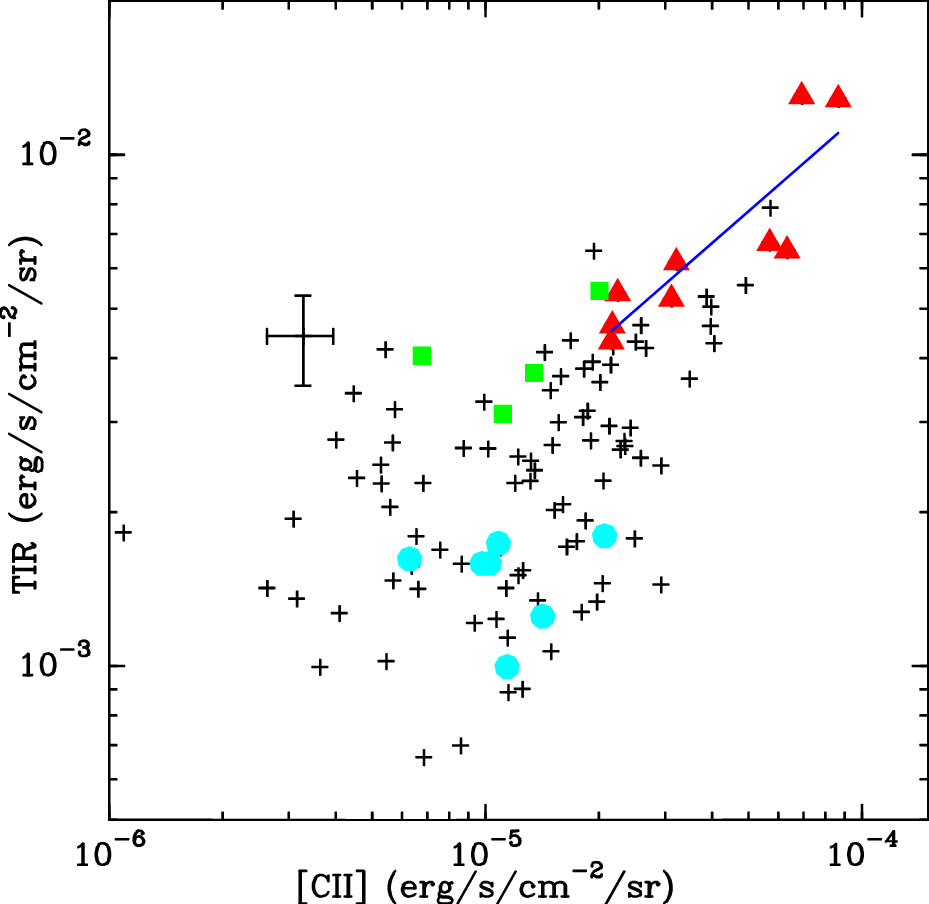}
\caption{ Correlation plot of \CII\ and Total Infrared (TIR) intensities at
a 12\arcsec\ resolution and on a 12\arcsec\ grid. Markers and the straight
line are the same as in Fig.~\ref{fig_allscats}.
\label{fig_cplustir} }
\end{figure}

%

\begin{figure}[h]
\centering
\includegraphics[height=7.0cm,angle=-90]{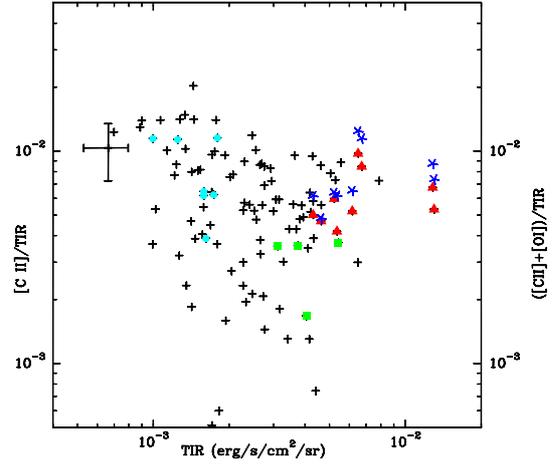}
\caption{ Plot of \CII/TIR and (\CII + \OI)/TIR (blue asterisks) in the
\HII\ region as a function of the
total infrared (TIR) intensities at a resolution of 12\arcsec\ on a
6\arcsec\ grid. The error bars denote 30\% errors. Markers same as in
Fig.~\ref{fig_allscats}.
\label{fig_heating}}
\end{figure}

\citet{boquien2010b} have obtained a linear fit to the total infrared
(TIR) luminosity as a function of the luminosity in the PACS
160$\,\mu$m\ band for the entire M\,33 galaxy. The TIR
  is the total infrared intensity, integrated between $1\,\mu$m and
  1\,mm. It is about a factor 2 \citep{rubin2009} larger than the FIR
  continuum which is integrated between 42.7 and 122\,$\mu$m.  
\citet{boquien2010b} derived the TIR from fits of \citet{draineli2007}
models to the MIPS, PACS, and SPIRE FIR data. And they find a tight
linear relation between the two quantities: ${\rm log} L_{\rm TIR} = a
\times {\rm log} L_{160} + b$ with $a$ = 1.013$\pm$0.008 and $b$=
0.429$\pm0.097$. We have used this relation to derive the TIR
intensity at each position of the mapped region, at a resolution of
12\arcsec.  To check that this is a valid approach, we independently
derived the TIR at individual positions by fitting a greybody to the
MIPS, PACS, and SPIRE data, smoothed to a common resolution of $40''$.
The resulting TIR agrees within 10\% with the TIR
derived from only the 160$\,\mu$m\ band.

Figure~\ref{fig_cplustir} shows a scatterplot between TIR and \CII.
Positions corresponding to the selected sub-regions are indicated
using different markers. We find that \CII\ and TIR are tightly
correlated with a correlation coefficient of 0.87 only in the \HII\
region ($A$) with the fitted slope being 0.64$\pm$0.14.  The
correlation coefficient in region $B$ is 0.52 and and $C$ shows no
correlation.
	
Incident FUV photons with energies high enough to eject electrons from
dust grains (h$\nu>6$ eV) heat the gas via these photoelectrons, with
a typical efficiency of 0.1--1\% \citep{hollenbach1997}.  Efficiency
is defined as the energy input to the gas divided by the total energy
of the FUV photons absorbed by dust grains. Based on ISO/LWS
observations of a sample of galaxies, \citet{malhotra2001} found that
(a) more than 60\% of the galaxies show $L_{\rm [C II]}/L_{\rm FIR}
>0.2$\% and (b) $L_{\rm [C II]}/L_{\rm FIR}$ decreases with warmer FIR
colors and increasing star formation activity, indicated by higher
$L_{\rm FIR}/L_{\rm B}$ ratios, where $L_{\rm B}$ is the luminosity in
the B band.  We have calculated the ratio of the intensities of
\CII/TIR, as a proxy for the heating efficiency, and plotted it
against with \CII\ intensities (Fig.~\ref{fig_heating}). The heating
efficiency vary by more than one order of magnitude within the
$2'\times2'$ region, between 0.07 and 1.5\%.  Considering an
uncertainty of 20\% in both the measured 160$\,\mu$m\ intensities and
the \CII\ intensities, for the relation between the 160$\,\mu$m\
luminosity and the TIR luminosity mentioned earlier, we estimate the
uncertainty in the \CII/TIR ratio to be $\sim 30$\%.  The observed
variation in the \CII/TIR intensity ratios is significantly larger
than the uncertainty we estimate from the \CII\ and 160$\,\mu$m\
intensities.  For the \HII\ region ($A$), we find heating efficiencies
between 0.2 and 1.0\%.  Regions $B$  and $C$ show efficiencies
between 0.15--0.4\% and 0.4--1.2\%.  Within the \HII\ region ($A$),
the total heating efficiency, including the \OI\ 63$\,\mu$m\ line,
(\CII +\OI )/TIR, lies between 0.3--1.2\% (Fig.~\ref{fig_heating}).
Outside the \HII\ region, reliable \OI\ data is largely missing.

\section{Modeling the PDR emission at the \Halpha\ peak of BCLMP 302}

\subsection{Estimate of FUV from TIR}

One of the key parameters of PDRs is the FUV radiation field which heats
the PDR. From energy considerations, the total infrared cooling emission
is a measure of the irradiating FUV photons of the embedded OB stars. We
estimate the FUV flux G$_0$ (6\,eV$<h\nu<$13.6\,eV) impinging onto the
cloud surfaces from the emergent total infrared intensities: G$_0$ =
4\,$\pi$ $I_{\rm FIR} = 4\,\pi\,0.5\,I_{\rm TIR}$ with G$_0$ in units of
the Habing field $1.6\,10^{-3}$\,erg\,s$^{-1}$\,cm$^{-2}$
\citep{habing1968} and the intensity of the far-infrared continuum
between $42.5\,\mu$m and $122.5\,\mu$m, $I_{\rm FIR}$, in units of
erg\,s$^{-1}$\,cm$^{-2}$\,sr$^{-1}$.  Here, we assume that heating by
photons with $\rm{h}\nu<6$\,eV contributes a factor of $\sim2$
\citep{TielensHollenbach1985} and that the bolometric dust continuum
intensity $I_{\rm TIR}$ is a factor of $\sim2$ larger than $I_{\rm FIR}$
\citep{dale2001}.

At the \Halpha\ peak position, the TIR intensity of 1.18$\times
10^{-2}$~erg~s$^{-1}$~cm$^{-2}$~sr$^{-1}$ translates into a FUV field of
$G_0=46$ in Habing units. Outside of the \Halpha\ peak, the FUV field,
estimated from the TIR, drops by more than one order of magnitude (cf.
Fig.\,\ref{fig_cplustir}).

The FUV radiation leaking out of the clouds is measured using the
GALEX UV data \citep{martin2005,gildepaz2007} to be
G$_0=24$ for a 12\arcsec\ aperture.

We thus find, that 66\% of the FUV photons emitted by the OB stars of
the \HII\ region are absorbed and re-radiated by the dust, and 34\%
are leaking out of the cloud, at the \Halpha\ position. This is
consistent with the FUV extinction derived from the \Halpha\ and
24$\,\mu$m\ fluxes \citep{relano2009} and the FUV/\Halpha\ reddening
curve \citep{calzetti2001}. 

It is interesting to note that \HII\ regions observed in the LMC and
other spiral galaxies by \citet{oey1997} and \citet{relano2002} show
that typically around 50\% of the ionizing stellar radiation escape
the \HII\ regions, roughly similar to the fraction of 36\% we find in
BCLMP\,302/M33.
%

\subsection{\CII\ emission from the ionized gas
\label{sec_cplus_ionized}}

\begin{table}
\begin{center}
  \caption{Properties of the \HII\ region BCLMP\,302}
  \label{tab_h2prop}
\begin{tabular}{lll}
\hline \hline
Excitation parameter $u$ & 180\,pc~cm$^{-2}$ & IK74 \\
Radius $r$ & 39 pc & IK74 \\
Mass & 10$^5$~\msun & IK74 \\
rms electron density & 6.2~\cmcub & IK74 \\
Electron density $n_e$ & 100~\cmcub & R08 \\
Ionization parameter $U$ & $-3.3$ & this paper \\
Effective temperature $T_{\rm eff}$ & 38,000\,K & this paper \\
H$\alpha$ luminosity & 2.2\,10$^{38}$\,erg\,s$^{-1}$ & KPNO map \\
\hline
\end{tabular}
\tablefoot{ As explained in the text, the value given for the
    electron density ($n_e$)is {\it assumed} to hold for BCLMP\,302.}
\tablebib{IK74~\citet[][]{israel1974}, R08~\citet[][]{rubin2008}}
\end{center} 
\end{table}

In Table~\ref{tab_h2prop} we present properties of the \HII\ region
BCLMP\,302 compiled from the literature and also calculated thereof.
These properties were used to estimate the fraction of \CII\ emission
contributed by this \HII\ region using the model calculations of
\citet{abel2005} and \citet{abel2006}. These authors have estimated
the \CII\ emission for a wide range of physical conditions in \HII\
regions, varying their electron density $n_e$, ionization parameter
$U$, and effective temperature $T_{\rm eff}$.

The \CII\ PACS map is centered on the bright \HII\ region BCLMP302
\citep{boulesteix1974}, which corresponds to the \HII\ region no.\,53
in the M33 catalog of \citet[][IK74]{israel1974}. Using the measured
radio flux and the formula $u ({\rm pc~cm}^{-2}) = 13.5 (S/{\rm
  f.u.})^{1/3} (D/{\rm kpc})^{2/3}$ \citep{israel1973}, IK74 estimate
the excitation parameter $u = 180$~pc~cm$^{-2}$ for a distance of
720~kpc.  IK74 also derive an rms electron density of 6.2\,\cmcub, and
estimate the radius $r$ of the \HII\ region to be 38.5\,pc. We did not
correct these results for the now better known distance, as it does
not affect our conclusions. For an electron temperature $T_e$ of
10,000~K, \citet{panagia1973} expresses the excitation parameter as $u
({\rm pc~cm} ^{-2}) = 2.2\times 10^{-19} \left
  [{N(L)}{(\beta-\beta_1)^{-1}} \right ]^{1/3}$, where
($\beta$-$\beta_1$) is the recombination rate to the excited levels of
hydrogen in units of cm$^3$~s$^{-1}$. Thus we get for the total flux
of ionizing photons $N(L)=4.6\,10^{43}\,u^3$~s$^{-1}$.  Using the
above values of the parameters, the dimensionless ionization parameter
$U$ was derived via $U= N(L)/(c\,4\,\pi\,r^2\,n_{\rm H+})$
\citep{evans1985,morisset2004}.

The rms electron density derived by IK74, provides only a lower limit
to the true electron density and observations of e.g.  the \SII\
doublet at 6754\,\AA\ are missing which would allow a direct estimate.
\citet{rubin2008} observed 25 \HII\ regions in M33 using {\it
Spitzer}-IRS, and concluded that their electron densities are
$\sim100$\,\cmcub.  Since there is no apparent reason for the $n_e$ in
BCLMP\,302 to be significantly different, we assume $n_e=100$\,\cmcub.
Hence, with $n_{\rm H+}=n_e$, we estimate $\log U=-3.3$
(Table\,\ref{tab_h2prop}).

The observed ratio of double and single ionized Nitrogen, \NIII\
57$\,\mu$m/\NII\ 122$\,\mu$m, at the \Halpha\ position
(Table\,\ref{tab_linerms}), indicates an effective temperature of the
ionizing stars of about 38,000\,K \citep{rubin1994}.

From the model calculations of \citet[][Fig.\,5]{abel2006}, we estimate
the fraction of \CII\ emission from the BCLMP\,302 \HII\ region.  The
\CII\ fraction increases with dropping electron density, but is only
weakly dependent on the ionization parameter and the stellar continuum
model.  Depending on the model, the resulting fraction varies slightly,
$\sim20-30$\%.  The fraction would drop to about 10\%, for an electron
density of $10^3$\,\cmcub.

Observations of the \NII\ 205$\,\mu$m\ line, in addition to the
122$\,\mu$m\ line would allow to estimate more accurately the electron
density of the \HII\ region.  The model calculations and observations
compiled by \citet{abel2006} show that the \CII\ intensities stemming
from \HII\ regions and the intensities of the \NII\ 205$\,\mu$m\ line
are tightly correlated.  \citet{abel2006} find
$\mathrm{Log}\left[I^{\mathrm{C^+}}_{\mathrm{H^+}}\right] = 0.937
\mathrm{Log}\left[I^{\mathrm{N II}}_{\mathrm{H^+}}\right] +
0.689$~(erg~cm$^{-2}$~s$^{-1}$). Assuming that 30\% of \CII\ emission
stems from the BCLMP\,302 \HII\ region at the \Halpha\ peak position,
we estimate a \NII\ 205$\,\mu$m\ intensity of $\sim 3.6
\times10^{-6}$~erg~s$^{-1}$ cm$^{-2}$~sr$^{-1}$ (cf.
Table\,\ref{tab_linerms}), corresponding to $3\times
10^{-18}$\,W\,m$^{-2}$. The expected \NII\ 205$\,\mu$m\ intensity is a
factor of 3 below the estimated rms of our PACS observations and hence
is consistent with the non-detection. Using the detected \NII\
122$\,\mu$m\ line, and the predicted \NII\ 205$\,\mu$m intensity from
above, the \NII\ 122$\,\mu$m/\NII\ 205$\,\mu$m ratio is 2.8 and this
corresponds to an $n_e$ of $10^2$~\cmcub\
\citep[][Fig.\,22]{abel2005}, which is consistent with the $n_e$
assumed by us for the above calculations.

\subsection{Comparison with PDR models}

\begin{figure}[h]
\centering
\includegraphics[width=7.0cm]{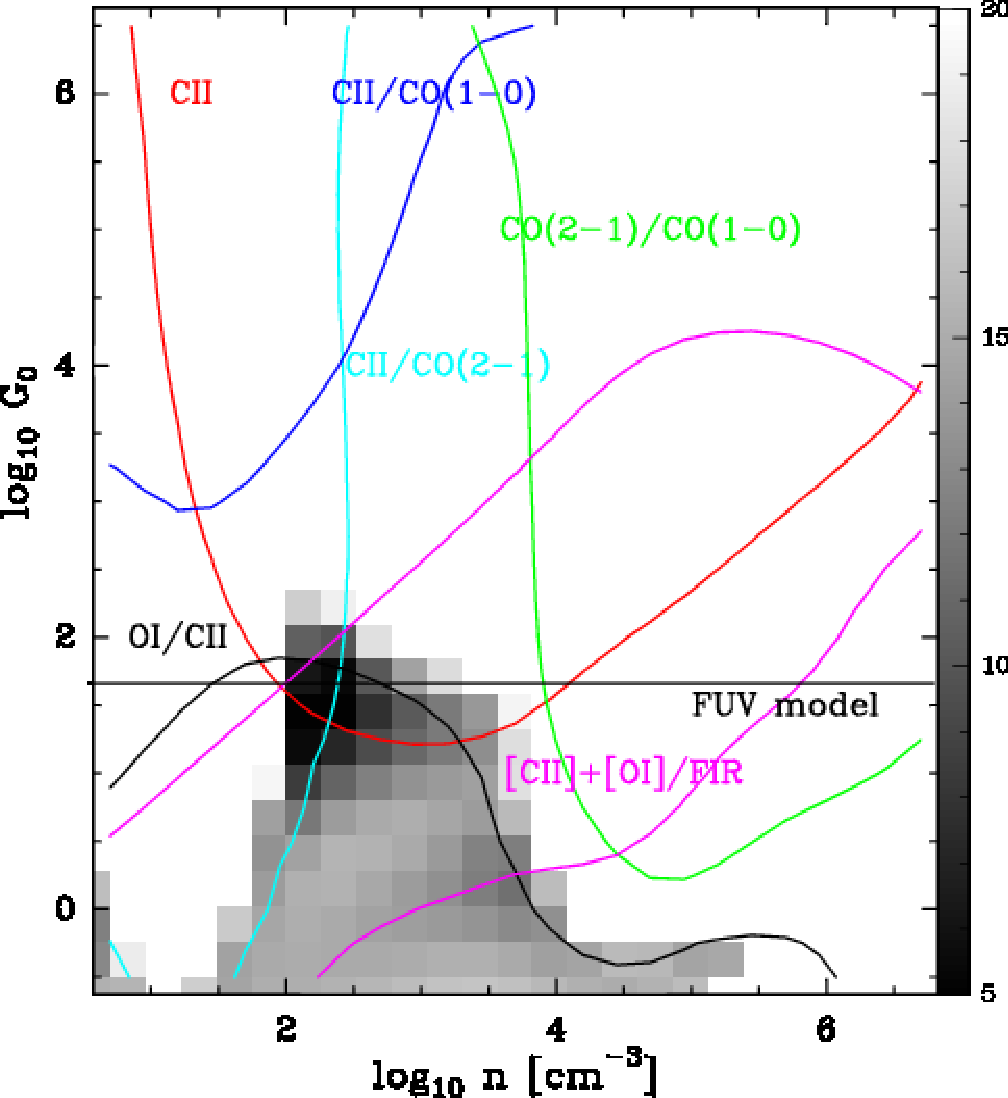}
\caption{Comparison of line intensities and intensity ratios at the
position of the \Halpha\ peak with plane-parallel
constant density PDR models \citep{Kaufman1999}. Grey-scales show the
estimated reduced $\chi^2$. The horizontal line shows the FUV
estimated from the total FIR intensity. The contours correspond to the
intensities/ratios of different spectral lines (as shown in the
labels).  The \CII\ intensity corresponds only to the estimated
contribution (70\% of the total) from the PDRs.
\label{fig_hifipdrmod}}
\end{figure}

\begin{figure}[h]
\centering
\includegraphics[width=7.0cm]{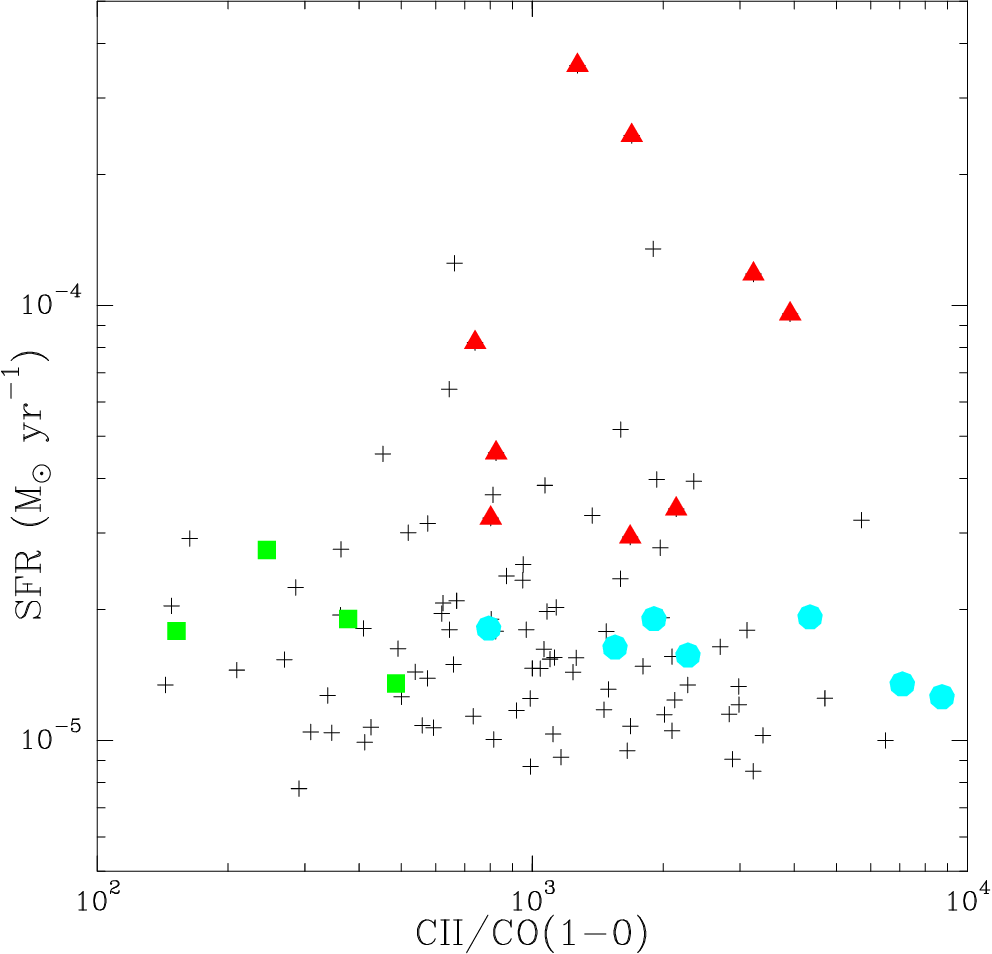}
\caption{Star formation rate (SFR) versus the \CII/CO(1--0) intensity
ratio. Markers are the same as in Figure~\ref{fig_allscats}.
  \label{fig_sfrvsciico}}
\end{figure}

\begin{table}
\begin{center}
  \caption{Input to the PDR model for the HIFI position
    \label{tab_intpdr}}
\begin{tabular}{lll}
\hline \hline
Tracer & Intensity \\ 
\hline
\CII$^\ast$ & 8.4~10$^{-5}$ \\
\OI  & 3.0~10$^{-5}$\\
CO (1--0)$^{\ast \ast}$ & 7.5~10$^{-9}$\\
CO (2--1) & 6.7~10$^{-8}$\\
TIR & 1.18~10$^{-2}$\\
\hline
\end{tabular}
\tablefoot{All intensities are in units of
  erg~sec$^{-1}$cm$^{-2}$~sr$^{-1}$. $^\ast$ The observed \CII\
  intensities was multiplied by 0.7, assuming that 30\% of the
  emission stems from the \HII\ region.  $^{\ast \ast}$ The CO 1--0
  data at 22\arcsec\ resolution were multiplied by the beamfilling
  factor 1.5, estimated by smoothing the 2--1 data from 11\arcsec\ to
  22\arcsec\ resolution. } 
\end{center} 
\end{table}

Here we investigate whether the FIR and millimeter lines, together
with the TIR, observed towards the \Halpha\ peak, can be consistently
explained in terms of  emission from the PDRs at the surfaces of the
molecular clouds.  We use the line and total infrared continuum
intensities at a common resolution of 12\arcsec.  
%

At the position of the \Halpha\ peak we observe the following
intensity ratios on the erg scale (cf.  Table\,\ref{tab_intpdr})
\OI/\CII\ = 0.35, \CII/CO(1--0) = 1.1$\times 10^4$, \CII/CO(2--1) =
1263, (\CII+\OI)/TIR = 9.6$\times 10^{-3}$ and CO(2--1)/CO(1--0) =
8.8. Comparing all ratios with the PDR model of \citet{Kaufman1999}
(Fig.\,\ref{fig_hifipdrmod}), we find a best fitting solution near a
FUV field of G$_0$=32 in units of the Habing field and a density of
320\,\cmcub. The fitted FUV field agrees rather well with the FUV
field of G$_0 = 46$ (in Habing units) estimated from the TIR
continuum. The corresponding reduced $\chi^2$ was estimated assuming
an error of 30\% on the observed intensity ratios.  
The reduced $\chi^2$ is defined as $\chi^2=1/(n-2)
  (\sum\limits_{i=0}^n {(I_i^{obs}-I_i^{model})^2}/\sigma_i^2$),
  where $(n-2)$ is the number of degrees of freedom with $n$ being the
  number of observed quantities used for the fitting and $I$ denotes
  either integrated intensities or ratios of integrated intensities.
We find the minimum value of $\chi^2$ to be 4.4, indicating that the
fit is not satisfactory.  Indeed, the fitted density seems rather low,
given the high critical densities of the \CII\ and in particular also
of the \OI\ line. On the other hand, the observed \OI/\CII\ ratio, is
consistent with the low density solution.  The ratios with the CO
lines, deviate from this solution. For instance, the CO 2--1/1--0
ratio together with the derived FUV, indicates a higher density of
about $10^4$\,\cmcub. Using line ratios to compare with the PDR model,
allows to ignore beam filling effects, to first order. Note that the
absolute \CII\ intensity, reduced by the fraction of 30\% stemming
from the \HII\ region, agrees well with the best fitting solution,
indicating a \CII\ beam filling factor of about 1.

The rather poor fit of the intensity ratios towards the H$\alpha$ peak
shows the short comings of a plane-parallel single density PDR model.
First tests using KOSMA-$\tau$ PDR models \citet{Roellig2006} of
spherical clumps with density gradients reconcile better to the
observations. This indicates strong density gradients along the line
of sight. In a second paper, we shall include new HIFI \CII\ data
along two cuts through the BCLMP\,302 region and explore detailed PDR
models, which include the effects of geometry and sub-solar
metallicity.

\section{Discussion \label{sec_discussion}}

Mapping observations of the northern inner arm of M\,33 at an
unprecedented spatial resolution of 50~pc have revealed details of the
distribution of the various components of the interstellar medium and
their contribution to the \CII\ emission. \CII\ at 158~\micron\
is one of the major cooling lines of the interstellar gas. Thus
irrespective of whether hydrogen is atomic or molecular, the \CII\ line
emission is expected to be strong wherever there is warm and
photodissociated gas. We have identified emission towards the \HII\
region as well as from the spiral arm seen in the continuum, and from a
region outside of the well-defined spiral arm.  \CII\ emission is
strongly correlated with the H$\alpha$ and dust continuum emission,
while there is little correlation with CO, and even less with \HI.  This
suggests that the cold neutral medium \citep[CNM;][]{Wolfire1995} does
not contribute significantly to the \CII\ emission in the BCLMP\,302
region. Recently, \citet{langer2010} found a similar poor correlation
between \CII\ and \HI\ emission in a sample of 29 diffuse clouds using
HIFI.   The lack of correlation between \CII\ and CO found in
BCLMP\,302, may indicate that significant parts of the molecular gas are
not traced by CO because it is photo-dissociated in the low-metallicity
environment of M33. This interpretation is consistent with both
theoretical models developed by \citet{bolatto1999} and recent
observational studies of diffuse clouds in the Milky Way by
\citet{langer2010}, and of dwarf galaxies by \citet{madden2011}.  

Comparison of the first velocity-resolved \CII\ spectrum of M33 (at the
\Halpha\ peak of BCLMP\,302) with CO line profiles show that the \CII\
profile is much broader, by a factor of $\sim1.6$, and slightly shifted
in velocity, by $\sim 1.6$~\kms.  Compared to the \HI\ line, at the same
angular resolution, the \CII\ line is less broad by a factor $\sim 1.3$
and shifted by $\sim 4.4$~\kms.  These findings indicate that the \CII\
line is not completely mixed with the CO emitting gas, but rather traces
a different more turbulent outer layer of gas with slightly different
systemic velocities, which is associated with the ionized gas.

%
%
  Interestingly, recent \CII\ HIFI observations of Galactic star
  forming regions \citep{Ossenkopf2010,Joblin2010} also show broadened
  and slightly shifted \CII\ line profiles relative to CO.
%

The two major cooling lines of PDRs are the \OI\ line at
63\,$\mu$m and the \CII\ 158\,$\mu$m line. The intensity ratios
[OI]/[CII] and ([OI]+[CII]) vs. the TIR continuum, have been used
extensively to estimate the density and FUV field of the emitting
regions. Using ISO/LWS, \citet{higdon2003} observed the far-infrared
spectra of the nucleus and six giant \HII\ regions in M\,33, not
including BCLMP\,302, but including NGC\,604, IC\,142, and NGC\,595
shown in Figure\,\ref{fig_overview}. The 70\arcsec\ ISO/LWS
beam corresponds to 285\,pc and therefore samples a mixture of the
different ISM phases. They find \OI/\CII\ line ratios in the range 0.7
to 1.3, similar to the range of values found in the center and spiral
arm positions of M83 and M51 \citep{kramer2005}. Towards the \HII\
region BCLMP\,302 in the northern inner arm of M\,33, we measure much
lower \OI/\CII\ ratios between 0.1--0.4.  These ratios lie towards the
lower end of the values found by \citet{malhotra2001} in their ISO/LWS
study of the unresolved emission of 60 galaxies who find values
between $\sim0.2$ and 2.  Similarly low values of down to 0.16 are
found e.g.  in the Galactic star forming regions DR\,21 and W3\,IRS5
\citep{jakob2007,kramer2004}. Comparison with the PDR models of
\citet[][Fig.\,4]{Kaufman1999} and \citet{Roellig2006} show that the
\OI\ 63$\,\mu$m line becomes stronger than the \CII\ emission in
regions of high densities of more than about $10^4$~\cmcub. At the
\Halpha\ peak position in BCLMP\,302, we observed a ratio of 0.4,
after correcting \CII\ for the contribution from the ionized gas.
This ratio indicates lower densities and a FUV field of less than
about 100\,G$_0$ (cf.\,Fig.\,\ref{fig_hifipdrmod}). Still lower
ratios, indicate lower impinging FUV fields.

The ratio of \OI\ +\CII\ emission over the FIR continuum, is a good
measure of the total cooling of the gas relative to the cooling of the
dust, reflecting the ratio of FUV energy heating the gas to the FUV
energy heating the grains, and hence the grain heating efficiency,
i.e. the efficiency of the photo-electric  (PE) effect
\citep{rubin2009}.  Efficiencies of up to about 5\% are still
consistent with FUV heating, i.e. with emission from PDRs
\citep{bakestielens1994,Kaufman1999}. The PE heating efficiency is a
function of FUV field, electron density, and temperature. A high FUV
field leads to a large fraction of ionized dust particles, lowering
the efficiency. On the other hand, low metallicities naturally lead to
increased efficiencies as e.g. the PDRs become larger when the dust
attenuation is lowered \citep{rubin2009}.

\citet{higdon2003} compared the \OI +\CII\ emission with the FIR(LWS)
continuum integrated between 43 and 197$\,\mu$m, and obtained ratios
of 0.2\% to 0.7\%, corresponding to 0.4 to 1.4\% for a rough estimate
of the FIR/TIR conversion factor of 2 \citep[][]{dale2001,rubin2009}.
Towards the $2'\times2'$ region presented here, the \CII/TIR ratio
varies by more than a factor 10 between 0.07 and 1.5\%.  Extra
galactic observations at resolutions of 1\,kpc or more, find
efficiencies of only up to $\sim0.3$\% \citep{kramer2005,
malhotra2001}. The observations of M\,31 by \citet{rodriguez2006} with
ISO/LWS at $\sim300$\,pc resolution, do also find high efficiencies of
up to 2\%.  \citet{rubin2009} analyzed BICE \CII\ maps of the LMC at
225\,pc resolution, and find efficiencies of upto 4\% in this low
metallicity environment. The mean values found in the LMC are however
much lower: 0.8\% in the diffuse regions, dropping to $\sim0.4$\% in
the SF regions. A similar variation was found in the Galactic plane
observations by \citet{nakagawa1998}.  Heating efficiencies observed
in the Milky Way span about two orders of magnitude.
\citet{habart2001} observed efficiencies as high as 3\% in the low-UV
irradiated Galactic PDR L1721, whereas \citet{Vastel2001} also using
ISO/LWS fluxes of \CII\ and \OI, found a very low heating efficiency
of 0.01\% in W49N.  The global value for the Milky Way from COBE
observations is $\sim0.15\%$ \citep{wright1991}. The galactic averages
observed by e.g.  \citet{malhotra2001}, are dominated by bright
emission from the nuclei where the SFR and the FUV fields are large,
hence lowering the photo-electric heating efficiencies.

Unresolved observations of external galaxies find a tight correlation
between CO and \CII\ emission. This was initially seen by
\citet{stacey1991} with the KAO, and subsequently supported by ISO/LWS
observations. However at spatial scales of 50~pc we do not detect such a
tight correlation in BCLMP\,302. This lack of correlation between CO and
\CII\ emission is already seen in the maps of spiral arms in M\,31 at
$\sim300$\,pc resolution \citep{rodriguez2006}.
As summarized by them, the \CII/CO(1--0) ratio varies from $\sim1300$ in
galactic disks to about 6000 in starbursts and to about 23,000 in the
LMC.  In BCLMP\,302/M\,33, we find that the \CII/CO(1--0) ratio varies
between 200 to 6000, with the \HII\ region having values between
800-5000 (Fig.\,\ref{fig_sfrvsciico}). The  region $C$ shows
large values of the \CII/CO(1--0) intensity ratios as CO is hardly
detected.  Thus, while the \CII/CO(1--0) intensity ratios are higher for
regions with high SFRs, there is no marked correlation between the two
quantities at scales below about 300\,pc. The tight correlation between
\CII\ and CO\ emission breaks down at scales resolving the spiral arms
of galaxies.

\end{document}